\documentclass[11pt,epsfig]{article}
\usepackage{moriond,epsfig}
\usepackage{graphicx}

\newcommand{\f}{\frac}

\def\Journal#1#2#3#4{{#1} {\bf #2}, #3 (#4)}

\def\NPB{{\em Nucl. Phys.} B}

\def\PRD{{\em Phys. Rev.} D}

\def\be{\begin{equation}}
\def\ee{\end{equation}}
\def\bea{\begin{eqnarray}}
\def\eea{\end{eqnarray}}

\begin{document}
\vspace*{4cm}
\title{THRESHOLD RESUMMED SPECTRA IN SEMI-INCLUSIVE B DECAYS}

\author{ G. RICCIARDI }

\address{Dipartimento di Scienze Fisiche, Universit\'a degli Studi di Napoli Federico II,
\\Complesso Universitario di Monte Sant'Angelo, Via Cintia, 80126
Napoli, Italy
\\ and I.N.F.N., Sezione di Napoli, Italy }

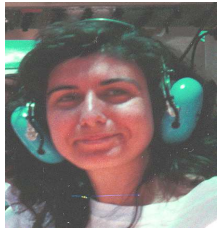
\maketitle\abstracts{I discuss some aspects of the universality of
soft gluon dynamics in semileptonic and radiative decays at the
threshold region.}

\section{Introduction}

Let us consider the threshold logarithmic resumming for
semi-inclusive B decays $B~\to~X_q +~\langle \mbox{~\scriptsize
leptons or~} \gamma \rangle$, where $ q \equiv u,\, d, \, s  $.
The perturbative calculation of such processes, at the threshold
region where $ E_X \gg m_X$, is plagued by large logarithms, which
originate
  from the incomplete cancellation of infrared real and virtual gluon
emissions. They can be factorized into the form
\begin{equation}
f(y) = \sum_{n=0}^{\infty}\sum_{k=0}^{2n} \, f_{n,k}\,
\alpha_S^n(Q)\, \log^k y~,\quad y=\f{Q^2}{m_X^2}
\end{equation}
where $Q$ is the hard scale of the process: at the threshold we
can set $Q= 2 \, E_X$. Both the logarithms and the argument of the
running coupling depend on the kinematics of the problem, by means
of the hard scale $Q$. The series contains at most two logarithms
for each power of $\alpha$, one of soft origin and another one of
collinear origin.
 The
terms of the
  series become large even if $\alpha_S\ll 1$:
 a truncated perturbative expansion is meaningless  and logarithmic
resummation is required in order to maintain a valid perturbative
expansion.

\section{Radiative and Semileptonic Decays}

Let us consider first the radiative decay with a real photon in
the final state
\begin{equation}
\label{bsgamma} B \, \rightarrow \, X_s \, + \, \gamma .
\end{equation}

The final hadronic energy is always large and it is of the order
of the heavy-flavor mass:
\begin{equation}
2 E_X \, = \, m_b \left( 1 \, + \, \frac{m_X^2}{m_b^2} \right) \,
\simeq \, m_b.
\end{equation}

Therefore,  $Q=m_b$ can be set and   the following resummation
formula holds for the invariant mass distribution:
\begin{equation}
\label{singola}
 \frac{ 1 }{ \Gamma_r } \frac{d\Gamma_r}{dt} \, = \,
C_r[\alpha_S(m_b)] \, \sigma[t;\alpha_S(m_b)] \, + \,
d_r[t;\,\alpha_S(m_b)]~,
\end{equation}
where $~t\equiv m_X^2/m_b^2~$. The coefficient $
C_r[\alpha_S(m_b)]$ and the remainder function
$d_r[t;\,\alpha_S(m_b)]$ are
  short-distance, process dependent functions.
  They have a reliable perturbative expansion in
  the QCD fixed coupling $\alpha_S(m_b)\simeq 0.22$.
 By definition, the remainder function does not contain large
logarithms in $t$. It behaves regularly at the threshold; that
means having, at the most, integrable singularities:
\begin{equation}
\lim_{t\to 0}\int_0^t d_r(t';\alpha_S)dt'=0
\end{equation}

The last term, $\sigma\left[t;\alpha_S(m_b)\right]$, is the
long-distance dominated, QCD form factor. It  factorizes the
threshold logarithms appearing in the perturbative expansion and
takes into account universal long-distance effects in radiative
and semileptonic decays.

Let us now consider the decay
\begin{equation}
\label{semilep} B \, \rightarrow \, X_u \, + \, l \, + \, \nu.
\end{equation}
It is possible to obtain a factorized form for the triple
differential distribution, the most general distribution in
process (\ref{semilep}) (its integration leads to all other
spectra):
\begin{equation}
\label{tripla} \frac{1}{\Gamma}\frac{d^3\Gamma}{dx du dw} \,=\,
C\left[x,w;\alpha(w\,m_b)\right]\,
\sigma\left[u;\alpha(w\,m_b)\right] \, + \,
d\left[x,u,w;\alpha(w\,m_b)\right],
\end{equation}
where:
\begin{equation}
w \,\equiv \, \frac{2 E_X}{m_b}~~~~~~~~(0\le w \le
2),~~~~~~~~~~~~~~~~~~~~~ x \,\equiv \, \frac{2
E_l}{m_b}~~~~~~~(0\le x \le 1)
\end{equation}
and
\begin{equation}
u \, \equiv  \, \frac{E_X - \sqrt{E_X^2 - m_X^2} }{E_X +
\sqrt{E_X^2 - m_X^2} } \, = \, \frac{1 - \sqrt{1 -
\left(2m_X/Q\right)^2} }{1 + \sqrt{1 - \left(2m_X/Q\right)^2} } \,
\simeq \, \left(\frac{m_X}{Q}\right)^2~~~~~~~~~~(0\le u \le 1).
\end{equation}

By passing from the two body radiative decay to the three body
semileptonic decay, the distribution acquires a dependence on the
charged lepton energy $E_l$. Moreover, and most important, while
in the radiative decay the hard scale $Q$ is always large (order
of $m_b$), this is no longer true for the semileptonic decay. We
have for instance the kinematical configuration with a large
invariant mass for the lepton pair, where $E_X$ is substantially
reduced. Therefore, we cannot set $Q=m_b$, as in the radiative
decay \cite{aglietti1,lav1,lav2,lav3}.

At this level, there is universality among radiative and
semi-leptonic decays, meaning that the same QCD form factor
$\sigma$ appears in both distributions (\ref{singola}) and
(\ref{tripla}), evaluated at the argument $u$ in the semileptonic
case and at $t$ in the radiative decay. The coupling constant
argument is set at the hard scale $Q=2 E_X$ in both processes; in
the radiative decay, that implies it is fixed to $m_b$.

A systematic logarithmic resummation is consistently done in
$N$-moment space or Mellin space. For example, the Mellin
transform of the radiative spectrum is of the form
\begin{equation}
\label{mellindist} \int_0^1 (1-t)^{N-1}
\frac{1}{\Gamma_r}\frac{d\Gamma_r}{d t} \, dt\,=\, C_r(\alpha_S)\,
\sigma_N(\alpha_S) \, + \, d_{r,N}(\alpha_S),
\end{equation}
where $ \sigma_N(\alpha_S) \, \equiv \, \int_0^1 dt \, (1-t)^{N-1}
\, \sigma(t;\,\alpha_S)$ and similarly for $d_{r,N}(\alpha_S)$.
The threshold region is studied in moment space by taking the
limit $N\rightarrow\infty$. It can  be shown
\cite{sterman,cattren1} that the form factor in $N$-space has the
following exponential structure:
\begin{equation}
\label{thresum} \sigma_N(\alpha_S) \, = \, e^{ G_N(\alpha_S) }.
\end{equation}
where $ G_N(\alpha_S) \, = \, \, l\, g_1(\lambda) + g_2(\lambda) +
\alpha_S \, g_3(\lambda)  + \alpha_S^2 \, g_4(\lambda)+ \cdots $
and $ \lambda\, \equiv \, \beta_0 \, \alpha_S \, \log N. $ The
$g_i(\lambda)$ are homogeneous functions of $\lambda$ and have a
series expansion around $\lambda = 0$. At LO (leading order)
approximation only the function $g_1$ is required, at NLO
(next-to-leading order) $g_1$ and $g_2$, at NNLO
(next-to-next-to-leading order) $g_3$ as well, and so on. The form
factor in $u$ space is recovered by the inverse Mellin transform.
It is convenient to define  the partially integrated or cumulative
form factor $\Sigma(u; \alpha_S)$:
\begin{equation}
\label{Sigmacomulativo}
 \Sigma(u;\alpha_S) \,=\,\int_0^u d
u^\prime \, \sigma(u^\prime;\,\alpha_S).
\end{equation}

To NNLO accuracy, one can write \cite{ugnnlo}:
\begin{equation}
\label{SigmaNNLO} \Sigma \left(u;\alpha_S\right) \,=\, \frac{  e^{
\, g_1(\tau) \, \log 1/u  +  \, g_2(\tau)  }  } { \Gamma\left[1 -
h_1(\tau) \right]  } \, \frac{S}{S|_{\tau \rightarrow 0}},
\end{equation}
where $ \tau \, = \, \beta_0 \, \alpha_S \,  \log 1/u $; by
definition $ h_1(\tau) \, \equiv \, \frac{d}{d\tau}\left[\tau
g_1(\tau)\right] \, = \, g_1(\tau) + \tau \, g_1'(\tau) $ and $$ S
= e^{ \alpha_S \, g_3(\tau) } \Bigg\{ 1 \, + \, \beta_0 \,
\alpha_S \, g_2'(\tau) \, \psi\left[1 - h_1(\tau) \right] \, +
\frac{1}{2} \beta_0 \, \alpha_S \, h_1'(\tau) \Big\{
\psi^2\left[1-h_1(\tau)\right] - \psi'\left[1-h_1(\tau)\right]
\Big\} \Bigg\}. $$
 The QCD cumulative form factor can be written in
an exponential form \cite{backtox}:
\begin{equation}
\label{expo}
 \Sigma  \, = \, e^G,
\end{equation}
where the expansion for the function $G$ is:
\begin{equation}
\label{expG} G(u;\alpha_S) \,=\, \sum_{n=1}^{\infty}
\sum_{k=1}^{n+1} G_{n k}\, \alpha_S^n \,  \log^k\frac{1}{u},
\end{equation}
and $G_{n k}$ are numerical coefficients.

\section{Single distributions for semileptonic decays}

All double and single distributions in semileptonic decays are
obtained by integrating the triple differential distribution
(\ref{tripla}). Some of these distributions may require also an
integration over the hard scale $Q$ up to $m_b$; this is different
from what happens in the radiative decays, where the hard scale is
fixed to $m_b$ . The decay spectra for (\ref{semilep}) can
therefore be divided into two classes~\cite{lav1,lav2,lav3}:
\begin{enumerate}
\item distributions in which the hadronic energy $E_X$ is not
integrated over. These distributions have the same
  infrared structure of the hadron invariant mass distribution of the
  radiative decay and
 can be related via
short-distance coefficients to the photon spectrum
(\ref{singola}). They share with the radiative decay the same
structure of threshold logarithms. In this sense, we recover
universality of long distance effects.
 \item distributions in which the
hadronic energy is integrated over and therefore all the hadronic
energies contribute. The distributions in this class have an
infrared structure different
  among themselves and from the hadron invariant mass distribution of the  radiative decay.
  The structure of the threshold logarithms is not the same of (\ref{singola}):
   there is no universality of long distance
effects.
\end{enumerate}

The single distribution in the hadron energy $E_X$ belongs to the
first class, while  the single distributions in $m_X$, in $E_l$
and in the light cone variable $p_+ = E_X - |\vec{p}_X|$
($\vec{p}_X$ defined as the three-momentum of the final hadron
state $X_q$) belong to the second group.

Let us consider, for instance,  the single distribution in the
hadron energy, obtained by integrating the triple differential
distribution in $u$ and $x$ \cite{lav1}:
\begin{equation}
\label{wg1impr} \frac{1}{2\Gamma}\frac{d\Gamma}{dw} =
C_{W1}\left(\alpha_S\right) \Big\{1 \, - \, C_{W2}
\left(\alpha_S\right)\, \Sigma\left[w-1;\,\alpha_S(m_b)\right] \,
+ H(w;\,\alpha_S) \Big\} \quad\!\! \quad\!\!(~w>1).
\end{equation}
$C_{W1}\left(\alpha_S\right)$ and $ C_{W2} \left(\alpha_S\right)$
are short distance coefficient functions and $H(w;\,\alpha_S)$ is
a remainder function, vanishing in $w=1$: they all have  standard
$\alpha_S$ expansions. We may  consider the parts of the spectrum
for $w<1$ and $w>1$ as two different spectra, merging in the point
$w=1$. The interesting case occurs at $w>1$, when resummation is
necessary; in fact, before the Sudakov shoulder, at $w <1$, there
are no large logarithms, and the spectrum  can be written as an
ordinary $\alpha_S$ expansion. The hadron energy distribution
contains $\Sigma$ defined in (\ref{Sigmacomulativo}), i.e. just
the integral of the form factor $\sigma$ entering the radiative
decay spectrum. The hadron energy spectrum is therefore directly
connected to the radiative decay.

This is not necessarily  true in  other kinds of semileptonic
spectra, as f.i. the distribution in the invariant hadron mass
squared, that is in the variable $ t \,=\, m_{X_q}^2/m_b^2  \,
\simeq \, u\, w^2 $. Such distribution is obtained by the triple
differential one (\ref{tripla}) by integrating in $x$ and
afterwards in  $u$ and $w$, with the appropriate kinematical
constraints. The integration in $w$ changes the logarithmic
structure of the distribution. It is still possible  to find a
resummed expression for the spectrum,  with  a form factor
factorizing large logarithms at all orders in perturbation theory,
but the form factor is now  effective, i.e. process dependent.
After the choice of a factorization scheme, we can write
\cite{lav2}:
\begin{equation}
\frac{1}{\Gamma}\frac{d\Gamma}{dt} \, = \, C_T(\alpha_S) \,
\sigma_T(t;\,\alpha_S) \, + \, d_T(t;\,\alpha_S),
\end{equation}
where:
\begin{equation}
\label{sigTdef} \sigma_T(t;\,\alpha_S) = \frac{\int_t^1
du/(2\sqrt{t \, u})\, C_H\left(\sqrt{t/u};\,\alpha_S(w \,
m_b)\right) \, \sigma\left[u;\,\alpha_S\left(m_b\sqrt{t/u}
\right)\right]} {\int_0^1 d w \, C_H(w;\,\alpha_S)}.
\end{equation}
$C_T(\alpha_S)$, $C_H(w;\,\alpha_S)$ are short distance
coefficient functions; $d_T(t;\,\alpha_S)$ is a remainder
function.

There is no simple connection between such semileptonic spectrum
and the hadron mass distribution in radiative decays. They have
different logarithmic structures. This can be explicitly checked
by building a cumulative form factor $\Sigma_T(t;\,\alpha_S)$
analogous to (\ref{Sigmacomulativo}). In a minimal factorization
scheme it is  also possible to build  the analogous of
exponentiated expression (\ref{expo}). In any case, by expanding
$\Sigma_T(t;\,\alpha_S)$ in a logarithmic series, and comparing
the coefficients of the logarithms with the coefficients $G_{nk}$
in ($\ref{expG}$), one finds that they  differ  already at
$O(\alpha_S)$ (in the single logarithm term).

\end{document}